\documentstyle[12pt]{article}
\begin{document}

 \begin{flushright}
            OCHA-PP-38\\
            NDA-FP-14\ \\
            December (1993)
 \end{flushright}
 \begin{center}
     {\LARGE  Swimming of Microorganisms \\
              Viewed from String and Membrane Theories}\\
      \vfill
    {\large Masako KAWAMURA,  Akio SUGAMOTO}\\
    \vspace{0.1in}
        {\it Department of Physics, Faculty of Science\\
         Ochanomizu University \\
        1-1 Otsuka 2, Bunkyo-ku, Tokyo 112, Japan}\\
        \vspace{0.2in}
        and \\
        \vspace{0.2in}
        {\large  Shin'ichi NOJIRI} \\
        \vspace{0.1in}
        {\it Department of Mathematics and Physics \\
        National Defence Academy \\
        Yokosuka, 239, Japan}
 \end{center}
 \vfill
\begin{abstract}
Swimming of microorganisms is studied from a viewpoint of extended objects
(strings and membranes) swimming in the incompressible fluid of low Reynolds
number. The flagellated motion is analyzed in two dimensional fluid, by using
the method developed in the ciliated motion with the Joukowski transformation.
Discussion is given on the conserved charges and the algebra which are
associated with the area (volume)- preserving diffeomorphisms giving the
swimming motion of microorganisms. It is also suggested that the $N$-point
string- and membrane-like amplitudes are useful for studying the collective
swimming motion of microorganisms when fluctuation of the vorticity
distribution exists in the sticky or rubber-like fluid.
\end{abstract}

\newpage

\def\mos{microorganism}
\def\vv{v_{\bar{z}}}

\section{Introduction}

In the high-energy physics, we usually study the world with the length scale of
$10^{-12} \sim 10^{-29} \mu$m, where field theories of particles as well as
strings and sometimes exotic membranes play the important roles in describing
theoretically the phenomena of this tiny world. Unfortunately, we have never
experimentally observed the dynamics of fission and fusion of strings and
membranes, except for the old-aged hadronic strings. A lot of theoretical
machineries have been so far developed to describe the dynamics of these
extended objects, examples of which are conformal field theories based on
Virasoro, Kac-Moody and W-algebras (area-preserving diffeomorphisms), the
field theory of strings and so on. Therefore, it is by no means a waste of
time to search for another region of physics where the knowledge of
high-energy physics will probably be useful.

In this paper, we take up the swimming problem of microorganisms as such an
interesting application, where we know the following fact on the ways of
swimming {\cite{nezumi}}:
There exist only three different universality classes of the swimming ways
of microorganisms; (1) Swimming with cilia is adopted by the spherical
organisms with the length scale of $20 \sim 2 \times 10^4 \mu$m, an example
of which is {\it paramecium}; (2) the smaller microorganisms with the size
of $1 \sim 50 \mu$m swim with flagella, an example of which is the
{\it sperm}; (3) the bacteria with the size of $0.2 \sim 5 \mu$m swim with
bacterial flagella the motion of which resembles the screwing of the
wine-opener. Why is it possible that such simple classification be realized
in the swimming problem of microorganisms?

Recently, Shapere and Wilczek {\cite{Wil}} have studied the swimming of
microorganisms from the gauge theoretical view point; Knowing one cycle
of the swimming motion gives a closed path on the space of shapes, they
have found that the net self-propulsion of translation and rotation in
the low Reynolds liquid are expressed in terms of the Wilson operator
on the closed path. Starting from their works, we will develop the
ciliated and flagellated motions and study on our problem from the
view point of algebraic structure existing in the deformation operation
of the microorganisms in the low Reynolds liquid, as well as from the
swimming dynamics of a group of mocroorganisms, which may be connected
to the $N$-point correlation function of string and membrane theories.

Let us recall the hydrodynamics. The Reynolds number $R$, giving the ratio
of the {\it kinetic term} over the {\it viscosity term} by $ R=\rho V L/\mu$,
where $\rho, V, L$, and $\mu$ denote density, typical velocity, typical
length scale and the coefficient of viscosity, respectively, satisfies
\begin{equation}
       R \leq 1 \times [L/{\rm mm}],    \label{1}
\end{equation}
for the velocity of microorganisms $V \leq 1$mm/s with $\rho=1{\rm g/cm^3}$
and $\mu=10^{-3}$Pas of water. Therefore, for the microorganisms with the
length $L \ll 1{\rm mm}$, we have $R \ll 1$ (the low Reynolds number fluid)
which leads to the following equations of motion for the incompressible fluid:
\begin{equation}
  \nabla \cdot  {\bf v} = 0,    \label{2}
\end{equation}
and
\begin{equation}
  \Delta {\bf v}  = \frac{1}{\mu} \nabla p,      \label{3.1}
\end{equation}
or equivalently
\begin{equation}
  \Delta (\nabla \times {\bf v}) = 0,        \label{3.2}
\end{equation}
where $p$ is the pressure and ${\bf v}(x)$ is the velocity field of the
fluid. The surface of a microorganism swimming in the $D=3$ dimensional
fluid of the real world forms a closed membrane at a fixed times $t$, the
position of which can be parametrized by introducing $(D-1)$ parameters
$\xi^i$ $(i=1, \cdots, D-1)$ as
$X^{\mu} = X^{\mu}(t; \xi^1, \cdots, \xi^{D-1})$. It is sometimes
instructive to consider the $D=2$ dimensional fluid. Then, the surface
of a ciliate (flagellate) becomes a closed (open) string and its position
can be described by a complex number
\[  Z = x^1 + ix^2 = Z(t; \theta),  \]
with $-\pi \leq \theta \leq \pi$. In the sticky fluid of $R=0$, there is
no slipping between the surface of a microorganisim  and the fluid, namely,
we have the matching condition
\begin{equation}
 {\bf v}({\bf x}={\bf X}(t; \xi))= \dot{{\bf X}}(t; \xi),    \label{4.1}
\end{equation}
or in the general coordinate system with the metric tensor $g^{\mu \nu}(x)$
\begin{equation}
   g^{\mu \nu}(x)v_{\mu}(x)|_{x = X(t; \xi)} = \dot{X}^{\nu}(t; \xi).
   \label{4.2}
\end{equation}

Now we will exemplify the ciliated as well as flagellated motion of
microorganisms in $D=2$ dimensional fluid. The ciliated motion was attacked
by Shapere and Wilczek several years ago, so that our main concern will be
the flagellate motion, but we first discuss the ciliated motion from a bit
different viewpoint by summarizing our notations.

\section{The ciliated motion in the D=2 fluid}
Time evolution of the envelope of the cilia can be viewed as a small but
time-dependent deformation of a unit circle in a properly chosen scale,
\begin{equation}
 Z(t, \theta) = s + \alpha(t, s),       \label{5}
\end{equation}
where $s=e^{i\theta}$ and $\alpha(t, s)$ is arbitrary temporally periodic
function with period $T$ satisfying $|\alpha(t, s)| \ll 1$ with
$-\pi \leq \theta \leq \pi$. The complex representation of the covariant
velocity vector $v_\mu$ can be denoted as
\begin{eqnarray}
2v_{\bar{z}}(z, \bar{z})& = &(v_1 +iv_2)(z, \bar{z})  \label{5.1}    \\
        2v_z(z, \bar{z})& = &(v_1 - iv_2)(z, \bar{z}),  \label{5.2}
\end{eqnarray}
since their transformation property under the conformal transformation
$z=f(w)$ reads
\begin{eqnarray}
v_{\bar{z}}(z, \bar{z}) & = &
\frac{\partial \bar{w}}{\partial \bar{z}}v_{\bar{w}}(w, \bar{w})
\label{6.1}       \\
v_z(z, \bar{z}) & = &
\frac{\partial w}{\partial z}v_w(w, \bar{w}).  \label{6.2}
\end{eqnarray}
The matching condition (\ref{4.2}) is then given by
\begin{equation}
2v_{\bar{z}} \left( s + \alpha (t, s),  s^{-1}
+ \overline{\alpha (t, s)} \right) =\dot{\alpha}(t, s).
       \label{7}
\end{equation}
The factors 2 in Eqs.(\ref{5.1}), (\ref{5.2}) and (\ref{7}) come from
the flat two dimensional metric expressed in terms of $z$ and
$\bar{z}$, that is, $g^{z\bar{z}}(z, \bar{z})=2$. Perturbatively,
Eq.(\ref{7}) can be solved as
\begin{equation}
  2v_{\bar{z}}(s) = \dot{\alpha}(t, s)
- 2 \left[ \alpha (t, s)\partial_z {v_{\bar{z}}}^{(0)}
+ \overline{\alpha (t, s)} \partial_{\bar{z}} {v_{\bar{z}}}^{(0)}
\right]_{z=s}
  \label{8}
\end{equation}
with the solution of the lowest order ${v_{\bar{z}}}^{(0)}(z, \bar{z})$
satisfying
\begin{equation}
   2{v_{\bar{z}}}^{(0)}(z, \bar{z})|_{z=s} = \dot{\alpha}(t, s).
   \label{9}
\end{equation}
The general solution of the velocity vector for the $R \approx 0$,
incompressible  liquid is already known; the equations of motion (\ref{2})
and (\ref{3.1}) are then
\begin{equation}
  \partial_z v_{\bar{z}} + \partial_{\bar{z}} v_z = 0
  \label{10}
\end{equation}
and
\begin{equation}
  4\partial_z \partial_{\bar{z}} v_{\bar{z}} =
\frac{1}{\mu} \partial_{\bar{z}}p  \label{11}
\end{equation}
or equivalently
\begin{equation}
  \partial_z \partial_{\bar{z}}(\partial_z v_{\bar{z}}
- \partial_{\bar{z}} v_z)= 0.    \label{11.1}
\end{equation}
Introducing the stream function $U(z, \bar{z})$, Eq.(\ref{10}) is solved
as $v_{\bar{z}}= \partial_{\bar{z}}U$ and $v_z = -\partial_z U$,
and Eq.(\ref{11.1}) in terms of $U$, $(\partial_z \partial_{\bar{z}})^2
U(z, \bar{z})= 0$, has a general solution with two arbitrary holomorphic
functions $\phi_1 (z)$ and $\phi_0 (z)$ as
\begin{equation}
  U(z, \bar{z}) = \bar{z} \phi_1 (z) - z\overline{\phi_1 (z)}
+ \overline{\phi_0 (z)} - \phi_0 (z).
  \label{12}
\end{equation}
Now, we have the general solution for $v_{\bar{z}}$ as
\begin{equation}
   v_{\bar{z}}(z, \bar{z}) = \phi_1 (z) - z \overline{\phi'_1(z)}
+ \overline{\phi_2 (z)},    \label{13}
\end{equation}
where $\phi_2 (z) = \phi'_0 (z)$. If we require the velocity field
$v_{\bar{z}}$ be finite at spacial infinity, we have the expansion,
\begin{equation}
  \phi_1 (z) = \sum _{k<0} a_k z^{k+1} \hspace{0.3in}
  {\rm and} \hspace{0.3in} \phi_2 (z) = \sum_{k<-1} b_k z^{k+1}.
\label{14}
\end{equation}
Following Shapere and Wilczek, self-propulsion of translation and
rotation against the fluid can be extracted as the counterflows of
$-a_{-1}$ and $-$Im$\overline{b_{-2}}$, since the translational and
rotational flows, $a_{-1}$ and Im$\overline{b_{-2}}$, surviving even at
spacial infinity after solving Eq.(\ref{4.1}) or (\ref{7}) should be
cancelled by the self-propulsion of the \mos. Here we will find the
following useful property; if the velocity vector is known on the unit
circle as $v(s)=\sum_{k=-\infty}^{+\infty} v_k s^{k+1}$, then the
velocity vector at any place can be easily determined as
\begin{equation}
 \vv (z, \bar{z})= v^{(-)}(z) + v^{(+)}(\bar{z})
+ ({\bar{z}}^{-1}-z)\overline{v'^{(-)}(z)},         \label{15}
\end{equation}
with the help of
\begin{eqnarray}
  v^{(+)}(\bar{z}) \equiv \sum_{k \geq 0} v_k {\bar{z}}^{-k-1}
\label{16.1} \\
  v^{(-)}(z) \equiv \sum_{k<0} v_k z^{k+1}.     \label{16.2}
\end{eqnarray}
Then from Eqs.(\ref{8}), (\ref{9}), and (\ref{15}), we obtain a general
expression of $\vv$ caused by the time dependent small deformation of
$\alpha (t, s)$ as
\begin{eqnarray}
 2\vv (s)& = & \dot{\alpha}(t, s)     \nonumber        \\
         &   & \mbox{} - \left\{ \alpha (t, s)
\left( \partial_s {\dot{\alpha}}^{(-)}(t, s)
- \overline{\partial_s {\dot{\alpha}}^{(-)}(t, s)} \right)
\right.         \nonumber   \\
         &   & \makebox[0.2in]{} + \left. \overline{\alpha (t, s)}
\left( \partial_{\bar{s}}{\dot{\alpha}}^{(+)}(t, \bar{s})
- s^2 \overline{\partial_s {\dot{\alpha}}^{(-)}(t, s)} \right)
\right\}.                        \label{17}
\end{eqnarray}
{}From this expression, the net translationally swimming velocity
$v_{T}^{(\rm cilia)}$ of the ciliated \mos\ reads
\begin{eqnarray}
    2v_{T}^{({\rm cilia})} & = & -\dot{\alpha}_0 (t)  \nonumber    \\
                     &  & \mbox{}+ \sum_{n \leq 1} n (\dot{\alpha}_n
\alpha_{-n+1} - \overline{\dot{\alpha}_n} \alpha_{n-1}
- \overline{\dot{\alpha}_n} \overline{\alpha_{-n+3}})   \nonumber  \\
                     &  & \mbox{} - \sum_{n>1} n \dot{\alpha _n}
\overline{\alpha _{n-1}},         \label{18}
\end{eqnarray}
where $\alpha_n (t)$ is defined by $\alpha (t, s) =
\sum_{n=-\infty}^{+\infty} \alpha_n (t) s^n$.
On the other hand, the net angular momentum $v_{R}^{({\rm cilia})}$
gained by the \mos\ from the fluid becomes
\begin{eqnarray}
 2v_{R}^{({\rm cilia})} & = &-{\rm Im} \left\{ \dot{\alpha}_1 (t)
\right.  \nonumber      \\
              &  & \makebox[0.3in]{} -
\sum_{n \leq 1} n (\dot{\alpha}_n \alpha_{-n+2}
- \overline{\dot{\alpha}_n} \alpha_{n}
- \overline{\dot{\alpha}_n} \overline{\alpha_{-n+2}})  \nonumber   \\
              &   & \makebox[0.3in]{}+ \left. \sum_{n>1} n \dot{\alpha _n}
\overline{\alpha _n} \right\}.                       \label{19}
\end{eqnarray}
The net translation and rotation resulted after the period $T$ come
from $O(\alpha^2)$ terms since the $O(\alpha)$ terms cancell after the
time integration over the period.

\section{The flagellated motion in D=2 fluid}
Microorganisms swimming using a single flagellum can be viewed as an
open string with two end points, H and T, where H and T represent the
head and the tail-end of a flagellum, respectively. Our discussion will
be given by assuming that the distance between H and T is time-independent
and is chosen to be 4 in a proper length scale. This assumption can be shown
to be valid for the flagellated motion by small deformations in the
incompressible fluid{\cite{masa}}. Then, at anytime $t$, we can take
a complex plane of $z$, where H and T are fixed on $z=2$ and $-2$,
respectively. This coordinate system $z$ can be viewed as that of the
space of {\it standard shapes} of Shapere and Wilczek. Time dependent,
but small deformation of the flagellate can be parametrized as
\begin{equation}
   Z(t, \theta) = 2(\cos \theta + i\sin \theta \alpha (t, \theta)) ,
\label{20}
\end{equation}
where the small deformation $\alpha (t, \theta)$ can be taken to be a real
number\footnote{When $\alpha$ is taken to be a complex number, the length
of the
flagellum is locally changeable at $O(\alpha)$.
For such an elastic flagellum, we have similar results to that of the ciliated
motion. In case of real $\alpha$, its length is locally preserved at
$O(\alpha)$, giving a non-elastic flagellum, which is the more realistic one.}
satisfying
\begin{equation}
  \alpha (t, \theta) = -\alpha (t, -\theta).        \label{21}
\end{equation}
Here, we parametrize the position of the flagellum twice, starting from
the end point T at $\theta = -\pi$, coming to the head H at $\theta = 0$,
and returning to T again at $\theta = \pi$. Motion of the two branches
corresponding to $-\pi \leq \theta \leq 0$ and $\pi \geq \theta \geq 0$
should move coincidentally, which requires the condition (\ref{21}). The
Joukowski transformation $z = z(w)= w + w^{-1}$, separates the two
coincident branches in the $z$ plane to form lower and upper parts of
an unit circle in the $w$ plane, outside domain of which we are able to
study the swimming problem of the flagellate in a quite similar fashion to
that of the ciliate,
The parametrization of our \mos\ in the
$w$ plane corresponding to Eq.(\ref{20}) is now
\begin{equation}
  W(t, \theta) = e^{i \theta}(1 + \alpha (t, \theta)) + O(\alpha^2).
       \label{22}
\end{equation}
The general solution of the velocity field $\vv$ in the $z$ plane can be
transferred to that of $v_{\bar{w}}$ in the $w$ plane, by using the
transformation property (\ref{6.1}), that is, we have generally
\begin{equation}
  v_{\bar{w}}(w, \bar{w}) = \psi_1 (w) - z(w)\frac{\partial \bar{w}}
{\partial \bar{z}} \overline{\psi'_1 (w)} + \overline{\psi_2 (w)},
\label{23}
\end{equation}
where the two arbitrary holomorphic functions have the following expansion
\begin{equation}
  \psi_1 (w) = \sum_{k<0} a_k w^{k+1} \hspace{0.3in} {\rm and}
\hspace{0.3in}     \psi_2 (w) = \sum_{k<-1} b_k w^{k+1}.            \label{24}
\end{equation}
In the study of the ciliated motion, the property (\ref{15}) was extremely
useful; the corresponding expression in the flagellated motion can also be
obtained from a given velocity vector on the unit circle in the $w$ plane
$v_{\bar{w}} = \sum_{k=-\infty}^{+\infty} v_k s^{k+1} $, by using the
general solution of (\ref{23}) and (\ref{24}). In terms of velocity field
of the real space, we have
\begin{eqnarray}
  \lefteqn{ \vv (w, \bar{w}) } \\
  &  & =(1 - w^2)^{-1} \left( \alpha + \beta w^{-1} -
                       w^{-2}  v^{(-)}(w) \right)    \nonumber         \\
  &  & + (1 - \bar{w}^{-2})^{-1} \left( \alpha \bar{w}^{-2}
+ \beta \bar{w}^{-1} + v^{(+)}(\bar{w}) \right)    \nonumber      \\
  &  & + (\bar{w} - \bar{w}^{-1})^{-3} \{ (w + w^{-1}) - (\bar{w}
+ \bar{w}^{-1}) \}       \nonumber      \\
  &  & \; \times \left\{ \bar{\beta}(\bar{w} + \bar{w}^{-1}) + 2 \bar{\alpha}
      -2 \overline{v^{(-)}(w)} + (\bar{w} - \bar{w}^{-1})
\overline{{v^{(-)}}'(w)}    \right\},
                   \label{25}
\end{eqnarray}
where $v^{(\pm)}$ are defined by
\begin{eqnarray}
  v^{(+)}(\bar{w}) \equiv \sum_{k \geq 0} v_k {\bar{w}}^{-k-1}  \label{zw} \\
  v^{(-)}(w) \equiv \sum_{k<0} v_k w^{k+1}.     \label{bzw}
\end{eqnarray}
The constants $\alpha$ and $\beta$ in Eq.(\ref{25}) are fixed so that
$\vv$ has no singularities on the special points of the head H and
tail-end T. From this requirement, we have
\begin{eqnarray}
    \alpha & = & \frac{1}{2} \left( v^{(-)}(1) + v^{(-)}(-1) \right)
= \sum_{l<0} v_{2l-1}
      \label{26.1}      \\
    \beta & = & \frac{1}{2} \left(v^{(-)}(1) - v^{(-)}(-1) \right)
= \sum_{l<0} v_{2l},             \label{26.2}
\end{eqnarray}
as well as
\begin{equation}
    v_{\bar{z}}(w, \bar{w})|_{w = \bar{w} = \pm1} = 0,       \label{27}
\end{equation}
which is quite consistent to our requirement of fixing the head and
tail-end position in $z$ and $w$ planes.

Next task is to solve the matching condition of (\ref{4.2}) which is
written in the $z$ plane as
\begin{equation}
    2\vv (W(t, \theta), \bar{W}(t, \theta)) = \dot{Z}(t, \theta).
    \label{28}
\end{equation}
Perturbatively, the solution of (\ref{28}) reads
\begin{eqnarray}
   2\vv (s) & = & (s - s^{-1}) \dot{\alpha}(t, \theta)   \nonumber    \\
            &  & \mbox{}- 2 \left[ s\alpha (t, \theta) \partial_w \vv^{(0)}
+ s^{-1} \overline{\alpha (t, \theta)} \partial_{\bar{w}} \vv^{(0)}
\right]|_{w = s},
            \label{29}
\end{eqnarray}
with the help of the lowest order solution $v_{\bar{w}}^{(0)}(w, \bar{w})$
satisfying
\begin{equation}
  2v_{\bar{w}}^{(0)}|_{w=s} = (1 - s^2)(s - s^{-1}) \dot{\alpha}(t, \theta)
  \label{30}
\end{equation}
on the unit circle ($s = e^{i\theta}$).
Using the mode expansion satisfying Eq.(\ref{21}),
\begin{equation}
  \alpha (t, \theta) = \sum_{n=1}^{\infty} \alpha_n (t) \sin n\theta,
  \label{31}
\end{equation}
the lowest order solution is obtained;
\begin{eqnarray}
 \lefteqn{ 2i\vv^{(0)}(w, \bar{w})} \\
  &  & = -\left\{ (w - w^{-1})\sum_{n \geq 1} \dot{\alpha}_n w^{-n}
+ (\bar{w} - \bar{w}^{-1})\sum_{n \geq 1} \dot{\alpha}_n \bar{w}^{-n} \right\}
  \nonumber     \\
  &  & \ \  + \{ (w + w^{-1}) - (\bar{w} + \bar{w}^{-1}) \}  \nonumber    \\
  &  & \ \  \times \left\{ \sum_{n \geq 1} n \dot{\alpha}_n \bar{w}^{-n}
- \frac{\bar{w} + \bar{w}^{-1}}{\bar{w} - \bar{w}^{-1}} \sum_{n \geq 1}
\dot{\alpha}_n \bar{w}^{-n}  \right\}.                 \label{32}
\end{eqnarray}
Then, from Eqs.(\ref{29})$\sim$(\ref{32}), we have derived the velocity
of the fluid which incorporates perturbatively the small but time-dependent
deformation of the flagellate as
\begin{eqnarray}
  \lefteqn{2\vv|_{w = e^{i\theta}}  =  2i \sin \theta
\sum_{n \geq 1}\dot{\alpha}_n(t) \sin n\theta }       \nonumber          \\
   &  &- 4 \sum_{n, m, \geq 1} \alpha_n(t) \dot{\alpha}_m(t)
\sin n\theta (\cos \theta \sin m\theta + m \sin \theta \cos m\theta ).
   \label{33}
\end{eqnarray}
Now, we are able to determine the net swimming velocity
${v_T}^{\rm(flagella)}$ gained by the flagellate motion of \mos s:
 \begin{equation}
  2v_T ^{(\rm flagella)} = -i \dot{\alpha}_1
  - \sum_{m \geq 1} m \alpha_m \dot{\alpha}_{m+1} + \sum_{m \geq 2} m
\alpha_m \dot{\alpha}_{m-1} ,   \label{34}
\end{equation}
On the other hand, the angular momentum ${v_R}^{(\rm flagella)}$ is given by
\begin{equation}
   2{v_R}^{(\rm flagella)}  = - \frac{1}{2} \dot{\alpha}_2.   \label{35}
\end{equation}
After the time integration over the period $T$, ${v_R}^{(\rm flagella)}$
vanishes since in our first order approximation, the length of the
flagellum is fixed in the incompressible fluid. Therefore the second order
approximation is necessary for the non-vanishing ${v_R}^{(\rm flagella)}$.

\section{The selection rules and the symmetry of microorganisms' swimming}

Even though the results in Eqns.(\ref{18}), (\ref{19}), (\ref{34}) and
(\ref{35}) are obtained perturbatively, we are able to read from them the
characteristics of the \mos s' swimming; In order for the ciliates to swim
or rotate, they need the coexistence of the two different Fourier modes of
$n_1$, $n_2$. The selection rules for the allowed ($n_1, n_2$) conbinations
are
\begin{eqnarray}
  i)& \hspace{0.3cm} |n_1 - n_2| =1
\hspace{0.3cm} {\rm or} \hspace{0.3cm} |n_1
+ n_2 - 2|=1
\hspace{0.5cm}& {\rm for\ the\
cililated\ translation} \\
 \label{36.1}
 ii) & \hspace{0.3cm} n_1=n_2 \hspace{0.3cm} {\rm or} \hspace{0.3cm} |n_1
+ n_2|=2    \hspace{1cm} & {\rm for\ the\ ciliated\ rotation},
   \label{36.2}
\end{eqnarray}
The corresponding selection rules for the flagellate motion are
\begin{equation}
   iii) \hspace{1cm} |n_1 - n_2| =1  \hspace{3cm}
  {\rm for\ the\ flagellate\ translation},
  \label{37.1}
\end{equation}
where the Fourier modes are $\sin n\theta$ in this case.

Viewing these selection rules, we are tempted to elucidate the algebraic
structure possibly existing in the background of the swimming mechanism.
It is similar to the Virasoro algebra, but is different from it. For
such a purpose, introduction of the {\lq\lq}action" will be convenient.
The {\lq\lq}action" $S$ reproducing the classical equations of motion of
the swimming of $N$ \mos s in the incompressible liquid with low Reynolds
number may be given by
\begin{eqnarray}
  S_N & = & \sum_{i=1}^{N} \int dt\, \int d^{D-1}\xi_{(i)}\, P_{\mu}^{(i)}
(t; \xi_{(i)}) \left[ \dot{X}^{\mu}_{(i)}(t; \xi_{(i)}) - v^{\mu}(X_{(i)}
(t; \xi_{(i)})) \right]    \nonumber     \\
    &   & \mbox{} +\frac{1}{2 \pi \alpha'} \int d^D x\, \sqrt{g(x)}\,
\left[ -\frac{1}{\mu}p(x) \partial_{\nu} v^{\nu}(x) + \frac{1}{4}
\omega_{\mu \nu}(x) \omega^{\mu \nu}(x) \right]     \label{38}
\end{eqnarray}
where the velocity field $\omega_{\mu \nu}(x)$ is given by
\begin{equation}
   \omega_{\mu \nu}(x) = \partial_{\mu} v_{\nu}
- \partial_{\nu}v_{\mu},           \label{39}
\end{equation}
whose $D=2$ expression is $\omega_{z, \bar{z}}(z, \bar{z})
= \partial_z v_{\bar{z}} - \partial_{\bar{z}}v_z$.

We have introduced the parameter $\alpha'$ so as to make $S$ dimensionless,
where $\alpha'$ has the dimension of $({\rm Length})^D({\rm Time})^{-2}$.
The reason why we have used the notation $\alpha'$, familiar in the string
theories to describe the Regge slope, will be understood later. The Lagrange
multiplier fields of $P_{\mu}^{(i)}(t; \xi_{(i)})$ $(i=1, \cdots, N$,
$\mu=1, \cdots, D)$ guarantee the matching condition of (\ref{4.1}) or
(\ref{4.2}) for $i$-th \mos\ at any time, and the pressure $p(x)$ is also
such multiplier giving the incompressibility given in Eq.(\ref{2}). The
field equation (\ref{3.1}) can be easily reproduced. In the action
(\ref{38}), time $t$ appears only in the first term of representing the
matching conditions, that is, the time evolution is triggerd only by the
self-motion of the \mos s, of which influence spreads instantaneously over
the whole space and causes the change of the fluid velocity there. Because
of the lacking of the kinetic term, we may call $S$ as the
 {\lq\lq}action". The additional metric contribution such as $\sqrt{g(x)}$
is only relevant for the curved space, an example of which has appeared
in the flagellate swimming on the $w$ plane. The later discussion is given
for the flat metric.

Now, we will define the following local transformation at a fixed time $t$:
\begin{eqnarray}
    \delta \dot{X}^{\mu}_{(i)} & = & \lambda^{\mu} (X_{(i)}
(t; \xi_{(i)}))                \label{40.1}      \\
    \delta P_{\mu}^{(i)} & = & 0           \label{40.2}    \\
    \delta v^{\mu} (x) & = & \lambda^{\mu}(x)       \label{40.3}    \\
    \delta p(x) & = & \kappa (x),          \label{40.4}
\end{eqnarray}
where we have assumed that the transformation parameters $\lambda^{\mu}(x)$
and $\kappa (x)$ are restricted by the equations of motions,
\begin{equation}
  \partial_{\mu} \lambda^{\mu}(x) = 0,  \hspace{0.2in} {\rm and}
\hspace{0.2in}  \partial_{\mu}{\lambda^{\mu}}_{\nu}(x)= \frac{1}{\mu}
\partial_{\nu} \kappa(x),   \label{41}
\end{equation}
where $\lambda_{\mu \nu} \equiv \partial_{\mu}\lambda_{\nu} -
\partial_{\nu}\lambda_{\mu}$ is the vorticity for $\lambda^{\mu}$.
Meaning of the transformations (\ref{40.1})$\sim$(\ref{40.4}) are quite
simple; the deformation of the \mos s (\ref{40.1}) triggers the increase
of the velocity field (\ref{40.3}) and of pressure (\ref{40.4}) so that
they can be consistent with the incompressible fluid dynamics of the low
Reynolds number. It is also important to note that the succession of these
time-independent transformations result in the time evolution of our
problem. Therefore, the transformations (\ref{40.1})$\sim$(\ref{40.4})
resemble the ordinary canonical transformation generated by the Hamiltonian.

Now, let us find the conserved current associated with the transformations
(\ref{40.1})$\sim$(\ref{40.4}). If we denote the Lagrangian of the fluid
itself as ${\cal L}_f$;
\begin{equation}
  {\cal L}_f = -\frac{1}{\mu} p \partial_{\mu} v^{\mu}
+ \frac{1}{4}\omega_{\mu \nu} \omega^{\mu \nu},         \label{42}
\end{equation}
then we have for the transformations (\ref{40.1})$\sim$(\ref{40.4}),
\begin{equation}
  \delta {\cal L}_f = \partial_{\mu} \left[ \left( \lambda^{\mu \nu}
- \frac{1}{\mu} g^{\mu \nu} \kappa \right) v_{\nu} \right].
  \label{43}
\end{equation}
Following the usual method, the conserved current associated with the
parameters $\lambda^{\mu}(x)$ and $\kappa (x)$ is obtained:
\begin{eqnarray}
  {J^{\mu}}_{[\kappa, \lambda^{\nu}]}(x)& = &
\frac{\delta {\cal L}_f}{\delta (\partial_{\mu}v_{\nu})} \delta v^{\nu}
- \left( \lambda^{\mu \nu}
- \frac{1}{\mu} g^{\mu \nu} \kappa \right) v_{\nu}       \nonumber       \\
  & = & \left( \omega^{\mu \nu} - \frac{1}{\mu} g^{\mu \nu} p \right)
\lambda_{\nu} - \left( \lambda^{\mu \nu} - \frac{1}{\mu} g^{\mu \nu}
\kappa \right) v_{\nu}.
    \label{44}
\end{eqnarray}
We should notice that the conservation of this current can be directly
proved with the help of the equations of motion (Eqs. (\ref{2}) and
(\ref{3.1})), namely,\begin{equation}
   \partial^{\mu} v_{\mu} = 0              \label{45.1}
\end{equation}
and
\begin{equation}
    \partial_{\mu} \left( \omega^{\mu \nu} - \frac{1}{\mu} g^{\mu \nu} p
\right) = 0,
    \label{45.2}
\end{equation}
as well as Eq.(\ref{41}). Instead of Eq.(\ref{45.2}), we can adopt the
conservation of the stress tensor
\begin{equation}
  \partial_{\mu} T^{\mu \nu} = 0,      \label{46}
\end{equation}
with
\begin{equation}
    T_{\mu \nu} = \mu (\partial_{\mu} v_{\nu} + \partial_{\nu} v_{\mu})
- g_{\mu \nu} p.            \label{47}
\end{equation}
Therefore, the conservation of the current
${J^{\mu}}_{[\kappa, \lambda^{\nu}]} (x)$ is essentially the reflection
of the incompressibility and the conservation of the stress tensor or the
balancing of the stress force of the fluid. Using the current conservation in
the outer region of the \mos s, Gauss theorem leads to a conservation law;
\begin{equation}
  \sum_{i=1}^{N} {Q_{(i)}}^{[\kappa, \lambda^{\nu}]} = 0,
   \label{48}
\end{equation}
where
\begin{equation}
    {Q_{(i)}}^{[\kappa, \lambda^{\nu}]} = \int_{S_{(i)}} d^{D-1}S^{\mu}\,
{J_{\mu}}^{[\kappa, \lambda^{\nu}]} (x)
    \label{49}
\end{equation}
is the integration over the surface $S_{(i)}$ of the $i$-th \mos.
It should be cautious about the flow at spacial infinity which corresponds
to the net translation and rotation of the group motion of the \mos s.
To take into account of this effect, it is convenient to include an
imaginary \mos, say $i=N$, at spacial infinity: $S_{(N)}$ may be an
infinity large envelope of the whole space, or in the compactified space
it can be an ordinary \mos\ located at the infinity point.

The transformation also generates the deformation of the shapes
of \mos\ (\ref{40.1}). The generator of this deformation per unit time can
be written as
\begin{equation}
  \hat{L}_{[\lambda^{\nu}]} \equiv \int d^{D-1} \xi \, \lambda^{\mu}
(X(t; \xi))\frac{\delta}{\delta X^{\mu}(t, \xi)},
  \label{50}
\end{equation}
which gives  \underline{the volume (area for $D=2$) preserving
diffeomorphisms} {\cite{Hop}} owing solely to the incompressibility
condition in Eq.(\ref{41}).
  The second condition in Eq.(\ref{41}) adds the further restriction on
$\hat{L}_{[\lambda^{\nu}]}$: By the help of the stream function
$\sigma_{\lambda}(x)$,
   the incompressibility  condition is automatically satisfied through
$\lambda^{\mu} = \epsilon^{\mu \nu \lambda}\partial_{\nu} \sigma_{\lambda}$,
   so that the Eq.(\ref{41}) becomes the constraint on the stream function
\begin{equation}
  \Delta (\Delta g_{\mu \nu} - \partial_{\mu}\partial_{\nu})
\sigma^{\nu}(x)=0,    \label{51.1}
\end{equation}
or in $D=2$, in terms of the only non-vanishing component $\sigma=\sigma_3$
\begin{equation}
  \Delta^2 \sigma (x)= 0.     \label{51.2}
\end{equation}
This constraint has been already solved generally in $D=3$ and $D=2$ fluid.

Here we will discuss a little more on the nature of the charge
$Q^{[\kappa, \lambda^{\nu}]}$ in Eq.(\ref{48}) as well as the algebraic
structure of the generator in Eq.(\ref{50}), taking up the simple case of D=2.
Then, we can use the knowledge given in Eqs.(\ref{10})$\sim$(\ref{13}).
The current $J^{\mu}(x)$ given in Eq.(\ref{44}) reads in D=2
\begin{equation}
   {J^z}_{[\kappa, \lambda]}(z, \bar{z}) = 2 \left( 2\omega_{\bar{z} z}
- \frac{1}{\mu}p \right) \lambda_{\bar{z}} - 2 \left( 2\lambda_{\bar{z}z}
- \frac{1}{\mu} \kappa \right) v_{\bar{z}},        \label{52}
\end{equation}
where the factors of 2 come from the metric $g^{z \bar{z}}=2$ needed for
raising and lowering the indices. The general expression of the parameters
$\kappa$ and $\lambda$ are obtained as
\begin{eqnarray}
  \lambda_{\bar{z}} & = & \epsilon_1(z) - z \overline{{\epsilon_1}'(z)}
+ \overline{\epsilon_2 (z)},     \label{53.1}    \\
  \lambda_{\bar{z} z} & = & 2(\overline{{\epsilon_1}'(z)} -  {\epsilon_1}'(z))
  \label{53.2}
\end{eqnarray}
and
\begin{equation}
   \kappa = -4 \mu ({\epsilon_1}'(z) + \overline{{\epsilon_1}'(z)}),
   \label{53.3}
\end{equation}
where we take the following most general expansion
\begin{equation}
  \epsilon_1 (z) = \bar{d}_L {\rm ln}z + \sum_{k=-\infty}^{+\infty} c_k
z^{k+1},   \label{54.1}
\end{equation}
and
\begin{equation}
   \epsilon_2 (z) = d_L {\rm ln}z + \sum_{k=-\infty}^{+\infty} d_k z^{k+1},
   \label{54.2}
\end{equation}
admitting the locally-defined but single-valued transformations.
The fields $\vv$, $\omega_{\bar{z} z}$, and $p$ have of course the
expressions similar to (\ref{53.1})$\sim$(\ref{53.3})'s, where
$\epsilon_1$ and $\epsilon_2$ are replaced by $\phi_1$ and $\phi_2$
in Eq.(\ref{14}). Now we have
\begin{equation}
  {J^z}_{[\epsilon_1, \epsilon_2]}(z, \bar{z}) = 16 \left[
\overline{{\phi_1}'}(\epsilon_1 + \overline{\epsilon_2}) -
\overline{{\epsilon_1}'}(\phi_1 + \overline{\phi_2}) \right].
    \label{55}
\end{equation}
The corresponding charge defined in Eq.(\ref{49}) on the circle, $|z|=r$,
takes the following form;
\begin{eqnarray}
  Q_{[\epsilon_1, \epsilon_2]} ^{(|z|=r)} & = &\frac{1}{2}
\left\{ \oint_{|z|=r} \frac{d \bar{z}}{2 \pi i} \, J^z +  \oint_{|z|=r}
\frac{dz}{2 \pi i} \, J^{\bar{z}}
  \right\}            \nonumber       \\
    & = & 8 \{ -(a_{-1} d_L + \overline{a_{-1}}
\overline{d_L})                               \nonumber  \\
    & + &  \sum_{k<-1} (k+1)(a_k d_{-k-2}
+ \overline{a_k} d_{-k         -2} +  b_k c_{-k-2}
+ \overline{b_k} \overline{c_{-k-2}}) \}.
    \label{56}
\end{eqnarray}
This means that there are infinite number of conserved charges which
are just equal to the coefficients $\{ a_k \}_{k<0}$ and $\{ b_k \}_{k<-1}$
of $\phi_1 (z)$ and $\phi_2 (z)$. In other words, two functions $\phi_1 (z)$
and $\phi_2 (z)$ give a solution of the fluid dynamics in the $D=2$ domain,
so that if $r$ is considered as the evolution parameter, the functional form
of them is conserved.

Next, we will study the algebraic structure of the operator in Eq.(\ref{50}).
Let's introduce the local version of its operator
\begin{equation}
  L_{[\lambda^{\nu}]} \equiv \lambda^{\mu}(x)\frac{\partial}{\partial x^{\mu}},
   \label{57}
\end{equation}
which becomes in $D=2$ as
\begin{eqnarray}
  L_{\sigma}& = &2 (\lambda_{\bar{z}} \partial_z
+ \lambda_z \partial_{\bar{z}})    \nonumber   \\
            & = & 2(\partial_{\bar{z}}\sigma \partial_z
- \partial_z \sigma \partial_{\bar{z}})
        \label{58}
\end{eqnarray}
when it is written in terms of the stream function $\sigma(z, \bar{z})$
having the follwing general form:
\begin{equation}
  \sigma(z, \bar{z}) = \bar{z}\epsilon_1 - z \overline{\epsilon_1 (z)}
              +\overline{\int^z dz \, \epsilon_2 (z)} - \int^z dz \,
\epsilon_2 (z).
    \label{59}
\end{equation}
Then, $L_{\sigma}$ can be considered as a Liouville operator of a dynamical
system moving in the phase space of $(z, \bar{z})$, having
$-\sigma(z, \bar{z})$ as its Hamiltonian. Invariance of the phase volume
during the temporal evolution of the dynamical system shows that $L_{\sigma}$
is really the area preserving diffeomorphism. The commutation relation is
simple, namely
\begin{equation}
  [L_{\sigma_1}, \:L_{\sigma_2}] = - L_{ \{ \sigma_1, \: \sigma_2 \} },
    \label{60}
\end{equation}
where the $\{ \sigma_1, \: \sigma_2 \}$ is the Poisson bracket defined by
\begin{equation}
   \{ \sigma_1, \: \sigma_2 \} = \partial_z \sigma_1 \partial_{\bar{z}}
\sigma_2 - \partial_{\bar{z}} \sigma_1 \partial_z \sigma_2.
   \label{61}
\end{equation}
The characteristic of our area-preserving diffeomorphism is summarized in
Eq.(\ref{59}) which can be expanded as
\begin{eqnarray}
  \sigma (z, \bar{z}) & = & \sum_{k} c_k \bar{z} z^{k+1} - \sum_{k}
\overline{c_k} z {\bar{z}}^{k+1}      \nonumber     \\
     &   & \; + \overline{d_{-2}} {\rm ln}\bar{z} - d_{-2}{\rm ln}z
        + \overline{d_L} \bar{z} ({\rm ln}z \bar{z} - 1 )
        - d_L z ({\rm ln}z \bar{z} -1)        \nonumber         \\
     &   & \; + \sum_{k\neq -2}\frac{1}{k+2} \overline{d_k}{\bar{z}}^{k+2}
-                    \sum_{k\neq -2}\frac{1}{k+2}d_k z^{k+2}.
     \label{62}
\end{eqnarray}
The generators corresponding to each term of the expansion may be denoted as
\begin{equation}
   L_{c_k} = L_{z^{k+1} \bar{z}}, \hspace{0.2in}
   L_{\overline{d_{-2}}} = L_{{\rm ln}\bar{z}}, \hspace{0.2in}
   {\rm and}    \hspace{0.2in}
   L_{d_L} = L_{z {\rm ln}z \bar{z} -1}, \hspace{0.12in} {\rm etc}..
   \label{63}
\end{equation}
Most of the generators are associated with the monomial
$z^{m_1}\bar{z}^{m_2}$ which can be denoted as $L_{(m_1, m_2)}$.
Then, the following commutation relation is easily understood.
\begin{equation}
  [L_{(m_1, \, m_2)}, \: L_{(n_1, \, n_2)}] = -(m_1n_2 - m_2 n_1)
L_{(m_1 + n_1 - 1, \, m_2 + n_2 - 1)},
    \label{64}
\end{equation}
which defines the so-called $W_{1 + \infty}$-algebra, having $L_{(0,\,0)}$
as its U(1) operator. The overall translations and rotation are generated
by $L_{c_{-1}}, \: L_{\bar{c}_{-1}} $ and $L_{b_{-2}} + L_{\bar{b}_{-2}}$,
respectively. Virasoro-like algebras are included among the generators;
\begin{eqnarray}
    \left[ L_{c_m}, \: L_{c_n} \right] &  = & -(m-n) L_{c_{m+n}},
\label{65.1}    \\
    \left[ L_{\bar{c}_m}, \: L_{\bar{c}_n} \right] & = & (m-n)
L_{\bar{c}_{m+n}},                   \label{65.2}
\end{eqnarray}
but
\begin{equation}
    [L_{c_m}, \: L_{\bar{c}_n}]  =  -(mn+ m+n) L_{(m+1, \, n+1)} \label{65.3}
\end{equation}
results in the appearance of a new generator $L_{(m+1, \, n+1)}$.
It is quite interesting to understand that net translation and rotation
for the ciliated \mos s are generated by the generator $L_{(m+1, \, n+1)}$,
where $|n \pm m| = 1$ corresponds to translation (\ref{36.1}) and $n=m$ or
$|n + m|= 2$ corresponds to rotation {\cite{masa}}.
As yet, the role of the generators $\{ L_{\sigma} \}$ in the swimming
problem is not manifestly clear but the algebra can depend on the shape of
the \mos s that really happens in the case of area-preserving diffeomorphisms
a part of which is included in our algebra. We could guess that the study on
the classification of possible area (volume)-preserving algebras of this
kind may lead to the understanding of three universality classes of \mos s'
swimming discussed in the beginning.

\section{Collective motion of \mos s and N-point correlation function}

Finally, starting from the {\lq\lq}action" (\ref{38}) of a group of \mos s,
we will study a possibility of using $N$-point correlation functions similar
to those of string and menbrane in the study of swimming motion in the $D=2$
and $D=3$ fluid, respectively. Consider the situation where the vortices are
created and annihilated, so that the probability of having the vortex
distribution (field) $\omega^{\mu \nu}(x)$ is given by
\begin{equation}
   P[\omega^{\mu \nu}(x)] \sim \exp \left\{
   -\frac{1}{2 \pi i \alpha'} \int d^D x \sqrt{g(x)} \: \frac{1}{4}
   \omega_{\mu \nu}(x) \omega^{\mu \nu} (x)     \right\},
   \label{66}
\end{equation}
where $i \alpha'$ is the external parameter controlling the fluctuation of
the vortex distribution. [The $\alpha' \rightarrow 0$ limit corresponds to
the classical limit without the fluctuation.] In this situation we should
sum over all the possible configurations of the velocity fields with
Eq.(\ref{66}) as their probability. The probability of having $N$ \mos s
with their surfaces located at $X_{(1)}, X_{(2)}, \cdots, X_{(N)}$, and
with their time derivatives
$\dot{X}_{(1)}, \dot{X}_{(2)}, \cdots,\dot{X}_{(N)}$, is given by the
following $N$-point correlation function (amplitude)
\begin{equation}
  G_N (X_{(1)},\dot{X}_{(1)}; \, X_{(2)},\dot{X}_{(2)}; \, \cdots; \,
X_{(N)}, \dot{X}_{(N)})  = Z_N/Z_0.
  \label{67}
\end{equation}
Here the partition function $Z_N$ is defined by
\begin{equation}
  Z_N = \int {\cal D} P_{\mu}^{(i)}(t,\xi_{(i)})\,
        \int {\cal D} p(x) \, \int {\cal D} v^{\mu}(x)\, \exp \{i S_N \},
  \label{68}
\end{equation}
with the {\lq\lq}action" $S_N$ in Eq.(\ref{38}). In the path integrations
over $P_{\mu}^{(i)}(t; \xi_{(i)})$ and $p(x)$ in Eq.(\ref{68}) guarantee
the matching condition (\ref{4.2}) and the incompressibility of the fluid,
respectively.        Moving to the momentum representation natually, we have
\begin{eqnarray}
   \lefteqn{ G_N (X_{(1)},\dot{X}_{(1)}; \, \cdots; \, X_{(N)},
\dot{X}_{(N)}) }      \nonumber        \\
   &  & = \int {\cal D} P_{\mu}^{(i)} \, \exp \left\{
          i \sum_{i=1}^{N} \int dt \, \int d^{D-1}\xi_{(i)} \,
P_{\mu}^{(i)}                \dot{X}^{\mu}  \right\}
          \nonumber     \\
   &  & \; \times  \: \tilde{G}_N ( X_{(1)}, P^{(1)}; \, \cdots \, ;X_{(N)},
P^{(N)} ),
      \label{69}
\end{eqnarray}
where
\begin{eqnarray}
   \lefteqn{  \tilde{G}_N ( X_{(1)}, P^{(1)}; \, \cdots \, ;X_{(N)}, P^{(N)} )}
   \nonumber     \\
   &   & = Z_0 ^{-1} \int {\cal D}v^{\mu}(x) \prod_{x} \delta (\partial v(x))
           \nonumber    \\
   &   & \; \times   \exp \left\{ -i\sum_{i=1}^{N} \int dt \,
\int d^{D-1}\xi_{(i)} P_{\mu}^{(i)} v^{\mu}(X_{(i)}(\xi_{(i)})) \right\}
          \nonumber    \\
    &    &  \; \times \exp \frac{1}{2 \pi i \alpha'}
           \int d^D x \, \frac{1}{2}v^{\mu} \Delta v_{\mu}
          \label{70}      \\
   &   &  = \exp \left[ 2 \pi i \alpha' \times \frac{1}{2}
          \sum_{i, j} \int dt_{(i)} \, \int d^{D-1}\xi_{(i)} \,
           \int dt_{(j)} \, \int d^{D-1}\xi_{(j)}  \right.
\nonumber    \\
    &   &  \; \times \left.  P_{\mu}^{(i)}(t_{(i)}; \xi_{(i)} )
          G_{\perp}^{\mu \nu} \left( X_{(i)}(\xi_{(i)}) - X_{(j)}(\xi_{(j)})
          \right)  P_{\mu}^{(j)}(t_{(j)}; \xi_{(j)} )  \right].
   \label{71}
\end{eqnarray}
Here the Green'n function $G_{\perp}^{\mu \nu}(x-x')$ satisfies
\begin{equation}
  \Delta G_{\perp}^{\mu \nu}(x-x') = \delta^{(D)}(x-x'),
    \label{72}
\end{equation}
and the transverse condition reflecting the incompressibility of the fluid,
namely
\begin{equation}
  \partial_{\mu} G_{\perp}^{\mu \nu}(x-x') = \partial'_{\nu}
  G_{\perp}^{\mu \nu}(x-x') = 0.
  \label{73}
\end{equation}
Now, we have obtained $N$-point correlation function for the collective
swimming of $N$ \mos s. It is quite similar to the $N$-point function of
strings {\cite{sakita}} for $D=2$ case and membranes {\cite{suga}} for $D=3$
case. It is also related to the string field theory, since the incoming and
outcoming strings are not point-like, but the Reggeons. More precisely,
refinement is necessary on the treatment of functional measure and boundary
condition in Eq.(\ref{70}). There is a possibility that the boundary
condition terms changes $\Delta g_{\mu \nu}$ into
$\Delta g_{\mu \nu} + S_{\nu} \partial_{\mu} - g_{\mu \nu} s^{\lambda}
\partial_{\lambda}$ with
\[ S_{\mu}(x)= \sum_{i=1}^{N} \int d^{D-1}S_{\mu}^{(i)} \,\delta^{(D)}
(x- X_{(i)}(t; \xi_{(i)})).   \]
It is also interesting to consider the multi-loop amplitudes if the meaning
of handle in the space of velocity field could be well understood.

Here we will discuss the use of $N$-point correlation function (\ref{69})
in the collective swimming motion of \mos s. If $G_N$ represents the
probability of having $N$ \mos s whose surfaces are located at $X_{(i)}$
with velocity $\dot{X}_{(i)}\ (i=1, \cdots, N)$, then it can be viewed as
the probability distribution of $\dot{X}_N$ of the imaginary \mos\ $N$
located at spacial infinity under the given data of $X_i$ and
$\dot{X}_i (i=1, \cdots, N-1)$. Following the usual strategy, the counterflow
$-\dot{X}_N$ can be indentified to the collective swimming motion of $N-1$
\mos s. Therefore, the averaged collective swimming motion over the
fluctuation distribution is given by
\begin{equation}
  -\langle \dot{X}_N \rangle = -\sum_{\dot{X}_N} \dot{X}_N
    G_N (X_1, \dot{X}_1; \cdots ; X_N, \dot{X}_N).
    \label{74}
\end{equation}
It is also interesting to study the response of the $N$-point correlation
function $G_N$, the Ward indentity, for the deformation of the shapes of
\mos s in Eq.(\ref{40.1})$\sim$(\ref{40.4})'s.

We hope much progress will come out along the strategies proposed in this
article.

\section*{Acknowledgments}

We are grateful to Shoichi Midorikawa for letting us know the interesting book
on the microorganisms' swimming. We also thank to Koichi Seo who has
introduced us the master thesis of S. Tanimura from which we know the
references on our problem. We give sincere gratitude to Ichiro Oda,
Kazuhiko Odaka for useful discussions.



\begin{thebibliography}{9}
\bibitem{nezumi}
  T. Motokawa, {\it Elephant Time and Mouse Time}
    (in Japanese), Chu\={o}-K\={o}ron (1992)
\bibitem{Wil}
  A. Shapere and F. Wilczek,  J.Fluid Mech. {\bf 198} (1989)557;\\
  S. Tanimura, Master thesis ``Tom and Berry -- on the geometrical
  aspect and the gauge theories of quantum and classical mechanics",
  Soryushiron kenkyu {\bf 85-1} (1992) 2 (in Japanese).
\bibitem{Hop}
  J. Hoppe, phD thesis ``Quantum theory of a massless relativistic
  surface and a two-dimensional bound state problem", Soryushiron Kenkyu
{\bf 80-3}  (1989) 145;\\
  D.B. Fairlie, P. Fletcher and C.K. Zachos,
Phys.Lett. {\bf 218B} (1989) 203;\\
  D.B. Fairlie and C.K. Zachos, Phys.Lett. {\bf 224B} (1989) 101;\\
  C.N. Pope and L.J. Romans, Phys.Lett. {\bf B226} (1989) 257.\\
  See also ``Proc. of the Trieste Conference Supermembranes and Physics in
  2+1 Dimensions (1989)" ed. by M.J. Duff, C.N. Pope, and E. Sezgin,
  World Scientific (1990).
\bibitem{Zamo}
  A.B. Zamolodochikov, TMP {\bf 65} (1985) 1205;\\
  V.A. Fateev and A.B. Zamolodochikov,
Nucl.Phys. {\bf B280} [FS18] (1987) 644;\\
   T. Homma, ``Area-preserving diffeomorphisms of the open membranes and the
  $W_{\infty}$ algebra", Science Univ. of Tokyo preprint (1992).
\bibitem{sakita}
  C.S. Hsue, B. Sakita and M.A. Virasoro, Phys.Rev. {\bf D2} (1970) 2857;\\
  J.-L. Gervais and B. Sakita, Phys.Rev. {\bf D4} (1971) 2291;\\
  S. Mandelstam, Nucl.Phys. {\bf B64} (1973) 205;\\
  A.M. Polyakov, Phys.Lett. {\bf 103B} (1981) 207, 211; \\
  For a review article, S. Mandelstam, Phys.Rep. {\bf 13C} (1974).
\bibitem{suga}
  A. Sugamoto, Nucl.Phys. {\bf B215} [FS7] (1983) 381.
\bibitem{masa}
  M. Kawamura, A. Sugamoto and S. Nojiri, in preparation

\end{thebibliography}
\end{document}